\documentclass[preprint]{vgtc}     

\usepackage{amsmath}
\usepackage{proof}%

\onlineid{1130}

\ieeedoi{10.1109/ISMAR67309.2025.00029}

\vgtccategory{Research}

\vgtcpapertype{evaluation}

\title{Comparing Hand and Controller Avatars with Hand Tracking and Controller-Based Interaction}

\author{\authororcid{Natalia Ocampo}{0009-0002-5929-9927}\thanks{e-mail:nataliaocampo@cmail.carleton.ca}\\ %
        \scriptsize Carleton University %
\and\authororcid{J. Felipe Gonzalez}{0000-0002-0716-1689}\thanks{e-mail: johannavila@cmail.carleton.ca}\\ %
     \scriptsize Carleton University %
\and \authororcid{Robert J. Teather}{0009-0007-4572-1820}\thanks{e-mail: rob.teather@monash.edu}\\ %
     \scriptsize Monash University
    }

\abstract{%
  Previous research suggests that the congruency between common VR input devices – such as controllers or hand tracking – and their visual representations (e.g., hand or controller avatars) influences user experience and performance. However, the specific effects of input-avatar combinations remain underexplored. We study the effects of common input devices (hand tracking and controllers) and visual representations (hand and controller avatars) on performance and perceived success in target acquisition tasks. We included both grasping and pinching gestures across 16 combinations of input, avatar, and target size. Results indicate that hand tracking benefits from any form of visual representation—even when mismatched—achieving up to 5.8\% greater accuracy compared to having no avatar, likely due to its reliance on visual feedback in the absence of a physical prop. Controllers were generally preferred and offered faster task completion. However, mismatched avatars had a stronger negative effect with controllers, particularly when the virtual gesture did not align with the physical action, leading to a 5.6\% drop in accuracy compared to the matched condition—suggesting that inaccurate feedback can be more disruptive than having no avatar feedback at all.

}

\keywords{Virtual Reality (VR), Hand Tracking, Handheld Controllers, Input Device, Avatar Representation, Gesture-Based Interaction, Target Acquisition}

\graphicspath{{figs/}{figures/}{pictures/}{images/}{./}} %

\usepackage{tabu}                      %
\usepackage{booktabs}                  %
\usepackage{lipsum}                    %
\usepackage{mwe}                       %
\usepackage{amsmath}                   %
\usepackage{mathptmx}                  %
\usepackage{svg}
\usepackage{graphicx}    %
\usepackage{subcaption}
\usepackage{placeins}

\begin{document}

\hyphenpenalty=9500  
\tolerance=1000

\maketitle

\section{Introduction}
Modern commercial virtual reality (VR) systems such as the Meta Quest \cite{meta_meta_2025}, HTC Vive \cite{htc_vive_2025} and PlayStation VR \cite{sony_playstation_2025} rely primarily on controllers as input devices to facilitate user interaction with the virtual environment (VE). Controllers provide precise tracking and the immediacy and reliability of button-based selection, making them well-suited to interactions requiring accuracy and control \cite{kangas_trade-off_2022, luong_controllers_2023, rantamaa_comparison_2023}. However, controllers require users to hold a physical object as a proxy for their real hand, despite offering considerably less natural interaction with objects in the environment. In this sense, controllers cause a disconnect between real-world hand movements and virtual actions, impacting user embodiment and presence \cite{wilson_violent_2018}.

Hand tracking, in contrast, is increasingly common as modern headsets \cite{meta_meta_2025} now integrate built-in cameras, so external devices such as the Leap Motion \cite{ultraleap_digital_2025} are no longer required. Hand tracking reduces the need for additional hardware and supports more natural gestures in VR, eliminating the need for physical controllers \cite{buckingham_hand_2021}. It benefits from greater similarity to real-world interaction by providing a much more direct correspondence between real hand to virtual hand movements, potentially improving user engagement through more intuitive interactions \cite{buckingham_hand_2021}. This is highly desirable in common VR applications such as gaming, education, training, healthcare, therapy, and physical fitness \cite{hamad_how_2022, hibbs_impact_2024, kim_yoga_2024}.

Although hand tracking enhances the sense of ownership \cite{adkins_evaluating_2021} and improves immersion \cite{kim_comparative_2022, luong_controllers_2023}, controllers remain the preferred input device for users. Controllers provide faster and more accurate interaction \cite{hameed_how_2023, hameed_evaluating_2021, luong_controllers_2023, masurovsky_controller-free_2020}, require lower overall workload in general tasks \cite{hameed_how_2023, kim_comparative_2022}, and foster greater user confidence in task execution \cite{kangas_trade-off_2022, pangestu_comparison_2022, voigt-antons_influence_2020}. A key factor that may contribute to these differences is the visual representation of the input device in VR. VEs typically provide users with a visual reference to represent their input method/device, typically a virtual hand and/or controller avatar \cite{adkins_evaluating_2021, argelaguet_role_2016}. Avatars are known to enhance immersion and embodiment by strengthening the connection between the user’s physical and virtual self \cite{hameed_evaluating_2021, kilteni_sense_2012, slater_influence_1998}. However, while maintaining a balance between physical and virtual interaction can improve engagement and task performance \cite{hibbs_impact_2024, kangas_trade-off_2022}, mismatched representations can have a negative impact \cite{luong_controllers_2023}.

VR platforms often face a trade-off between input devices (e.g., hand tracking or controllers) and visual representations (e.g., hand or controller avatars). For example, many applications rely on controller input for better accuracy, yet display a hand avatar to enhance embodiment, despite the mismatched representation. Other applications use hand tracking with matching hand avatars to support natural interaction in scenarios where performance is less critical. Many VR games mismatch visual representations entirely, presenting hands or in-game tools with either controller or hand-based input. The extent to which performance and user experience are influenced by the matching or mismatching of input to visual representation remains poorly understood. Existing research has typically focused on a limited number of input modality combinations, visual representations, and gestures \cite{johnson_exploring_2023, venkatakrishnan_how_2023}. A more comprehensive understanding of how visual representation interacts with different input devices is still needed to guide VR design decisions.

We present a study evaluating user preferences and performance under varying combinations of input device and avatar representation. We aim to address the question: \textit{\textbf{how do different input devices and visual representations influence user experience and performance?}} We recruited 48 participants who completed a series of object manipulation tasks in a custom VR game. The task was modeled after the standardized Fitts’ law selection paradigm \cite{clark_extending_2020}, and employed a methodology similar to previous studies \cite{kourtesis_action-specific_2022, monteiro_exploring_2023}. Participants performed tasks using two common input methods—hand tracking and controllers—executing two typical gestures: pinch and grasp. Visual representation conditions varied based on the presence or absence of a hand avatar and a controller avatar.

To our knowledge, this is the first study to systematically evaluate the impact of avatar representation of the input device across both hand tracking and controller input. While previous work has explored related phenomena \cite{johnson_exploring_2023, venkatakrishnan_how_2023}, they often merge input modality and avatar effects, lacking a controlled and systematic comparison—making it impossible to differentiate their individual effects. The main contribution of our paper is a rigorous and systematic analysis of the effects of input devices (hand tracking and controller) across various combinations of avatar representations (hand and controller avatars) on both task performance and user experience.

\section{Related Work}

\subsection{Controllers}

Controllers have been the de facto standard VR input device for decades due to their precise tracking, passive haptics, embedded vibrotactile feedback, and long-standing familiarity from consistent use in gaming and other interactive systems \cite{buckingham_hand_2021, masurovsky_controller-free_2020, novacek_overview_2020, rantamaa_comparison_2023}. These attributes are why they continue to be used in the most current commercial VR platforms, such as Meta Quest \cite{meta_meta_2025} and HTC Vive \cite{htc_vive_2025}. As controllers often use motion sensors, infrared tracking, and pressure sensitive buttons to detect user input and translate them into virtual actions, they offer higher accuracy \cite{kangas_trade-off_2022}, stability, and responsiveness in object manipulation tasks\cite{lin_need_2016, luong_controllers_2023, ponton_stretch_2024}. Moreover, the use of vibrotactile motors enhance user interaction by providing physical sensations that help with gestures like grasping, pressing, or triggering virtual objects \cite{cabaret_does_2024, wee_haptic_2021}. They enable more efficient task completion and greater precision in selection and trajectory-tracing tasks compared to input methods like hand tracking or mouse and keyboard \cite{kangas_trade-off_2022}.

Despite these benefits, controllers offer limited embodiment and immersion. Since controllers do not support natural mappings between real hand gestures and virtual actions, users are required to heavily rely on button presses. This lowers the sense of presence and agency in VR experiences \cite{wilson_violent_2018}. Previous studies \cite{lin_need_2016, ponton_stretch_2024} highlight that controllers may negatively impact body ownership due to the misalignment between a user’s real hand and their virtual hand avatar. In addition to impacting immersion and (indirectly) presence, it also increases cognitive load, as controller interactions feel less intuitive compared to natural hand movements. Additionally, traditional controllers have difficulty facilitating natural interactions due to their inability to support five-finger gestures. At least two fingers are used to grip the controller against the palm, leaving the remaining fingers for button interactions. Nevertheless, individuals may require more fingers for controller support or struggle to perform button interactions due to their hand size. Previous studies \cite{brown_evaluating_2013} have also highlighted the challenges individuals with smaller hands encounter compared to those with larger hands.

\subsection{Hand Tracking}

Hand tracking allows users to interact with VEs using hand gestures. Instead of relying on physical input (e.g., buttons), hand tracking employs computer vision and sensors to detect and translate hand movements into virtual actions \cite{buckingham_hand_2021}. Compared to controllers, hand tracking offers greater realism as users can engage with the virtual world using a 1:1 mapping between their real hand and virtual hand gestures, without the need to learn controller mappings. This improves embodiment as users perceive their virtual hands as an extension of their real hands. Johnson et al. \cite{johnson_exploring_2023} found that participants manipulating virtual objects when using hand tracking reported higher levels of perceived naturalness compared to controllers. Similarly, Argelaguet et al. \cite{argelaguet_role_2016} reported users preferred free-hand interactions for mid-air tasks as it offered a more direct and engaging experience.

However, hand tracking faces several challenges in tracking fidelity and accuracy. Unlike controllers, which provide stable tracking, hand tracking systems are more susceptible to occlusion, latency, and inconsistent gesture recognition, all of which reduce performance in precision tasks \cite{kangas_trade-off_2022}. The absence of tactile feedback with hand tracking also reduces the sense of realism. Even simple vibration offered by controllers provides additional tactile feedback missing with hand tracking. Without it, hand tracking systems lack tactile confirmation when interacting with virtual objects, negatively impacting usability, task performance \cite{masurovsky_controller-free_2020}, and user preferences compared to other alternatives \cite{johnson_exploring_2023}.

\subsection{Visual Representation}

Visual representations in VR play an important role in maintaining user engagement and improving interaction fidelity. Avatars representing the user’s hands or controllers provide a connection between physical actions and virtual responses. The accuracy of the avatar (in terms of appearance, alignment to the real hand, etc.) affect immersion, realism, and user performance \cite{zhang_understanding_2023}. For instance, Lin et al. \cite{lin_need_2016} report that appearance and behavior of virtual hands significantly influence user perception and control. Users exhibited stronger embodiment when virtual hands closely resembled their real hands in shape and motion. Similar effects have been reported in the use of avatars for controllers. Ponton et al. \cite{ponton_stretch_2024} investigated the effects of controller-avatar alignment, showing that maintaining congruency between the physical controller and the virtual counterpart improves performance and embodiment. Other research \cite{hibbs_impact_2024, ponton_stretch_2024} suggests congruency between physical actions and visual feedback reduces discomfort and improves task performance. Hibbs et al. \cite{hibbs_impact_2024} studied the effects of visual-physical synchronization in a VR cycling experiment. Their findings suggest that accurate alignment between user movements and avatar actions enhances realism, body ownership, and overall performance. Hanashima et al. \cite{hanashima_how_2022} similarly demonstrated that visuo-motor-tactile synchrony improves embodiment.

Perhaps the most similar study to ours, Venkatakrishnan et al. \cite{venkatakrishnan_how_2023} explored the interaction between input method and visual representation by evaluating three combinations: controller with a controller avatar, controller with a hand avatar, and glove-based hand tracking with a hand avatar, in a task involving moving doors. Similarly, Johnson et al. \cite{johnson_exploring_2023} investigated this relationship in a ball-sorting task using a pinching gesture. Their study included one hand tracking condition (with a hand avatar) and four controller-based conditions with different avatar combinations. While these studies offer valuable early insights into the impact of input and avatar combinations, they are limited in scope. Specifically, they do not systematically evaluate the full range of possible matching and mismatching combinations between input methods and visual representations. In both cases, the input method is tightly coupled with the avatar type, restricting the ability to isolate their individual effects. Additionally, the analysis is constrained to a narrow set of performance metrics and gesture types. Building on this past work, we aim to provide a more comprehensive evaluation by systematically varying both input method and avatar visual representations. We extend prior work by incorporating multiple gestures and a broader set of objective and subjective performance metrics to better understand how these factors influence user experience and task performance.

\section{Methodology}
Our study explores combinations of input devices across visual representations, with two common interaction gestures (grasping and pinching). We developed a VR game inspired by the Fitts’ law \cite{clark_extending_2020} reciprocal selection tasks used extensively as a standard \cite{soukoreff_towards_2004} to evaluate pointing devices in HCI and VR \cite{kourtesis_action-specific_2022, monteiro_exploring_2023}.

\subsection{Participants}

Via posters at our university and through social media, we recruited a diverse cohort of 48 participants, aged 18 to 30 years (mean age of 20.1 years), 27 self-identified men (56.3\%), 18 self-identified women (37.5\%), and 3 self-identified non-binary (6.3\%). Concerning visual acuity, all but 2 participants had normal (33 participants, 68.6\%) to corrected-to-normal (13 participants, 27.1\%) vision. Most (42, 87.5\%) were right-handed, while 5 (10.4\%) were left-handed, and 1 (2.1\%) was ambidextrous. VR experience varied widely: 27 participants (56.3\%) had limited exposure (only using VR a few times); 15 participants (31.3\%) were novice users; and 6 participants (12.5\%) had no prior VR experience. Additionally, 36 (75\%) had never used hand tracking, while 11 (22.9\%) had limited hand tracking experience. A single participant (2.1\%) reported extensive hand tracking experience. Participants had diverse levels of video game experience, self-reported as playing every day (9, or 18.8\%), weekly (8, or 16.7\%), occasionally on a monthly basis (11, or 22.9\%), yearly (10, or 20.8\%), and once a year or almost never (5, or 10.4\%).

\subsection{Apparatus}
\subsubsection{Hardware}

We used a Meta Quest 3 head-mounted display, which features a field of view of 110 degrees horizontally and 96 degrees vertically, a display resolution of 2064 × 2208 pixels per eye, and a Snapdragon XR2 Gen 2 processor for standalone operation. Notably, it also includes an integrated hand tracking system and two controllers that were used as input devices in the experiment.

Depending on the experimental condition and task, participants used either Quest 3’s controller or hand tracking capabilities to perform two different interaction gestures: the grasp gesture and the pinch gesture. Each gesture was performed similarly between input devices, subject to their operational differences. The grasp gesture required participants to clench their hand into a fist with hand tracking, or simultaneously press the {\rmfamily\textsc{Grip}}, {\rmfamily\textsc{{Trigger}}, and {\rmfamily\textsc{A}} buttons when using the controller. To perform the pinch gesture with hand tracking, participants would bring together the tips of their index finger and thumb or would simultaneously press the {\rmfamily\textsc{{Trigger}} and {\rmfamily\textsc{A}} buttons when using the controller.

\subsubsection{Software}

We used Unity 2022.3.1f1 and the Meta XR All-in-One SDK to develop the VR software used in the study. The VE was retrieved from the XR All-in-One SDK. It included an open, minimalist virtual space with a few low-detail items placed within the user’s vicinity designed to create a focused environment with minimal distractions. The camera was initially positioned at the center of the room, 3 m above the floor, oriented along the positive x-axis in world space. This setup helped prevent height misalignment between the user and the evaluation environment when entering the virtual world. The evaluation environment was a static planting bed\footnote{The planting bed model used for the VE was retrieved from Sketchfab at \href{https://skfb.ly/oGGpG.}{https://skfb.ly/oGGpG.}} placed directly in front of the camera and presented pre-located interactive buttons, tasks, and instructions. See Figure \ref{fig:planting-bed}

\begin{figure}[ht]
  \centering
  \includegraphics[width=0.8\columnwidth]{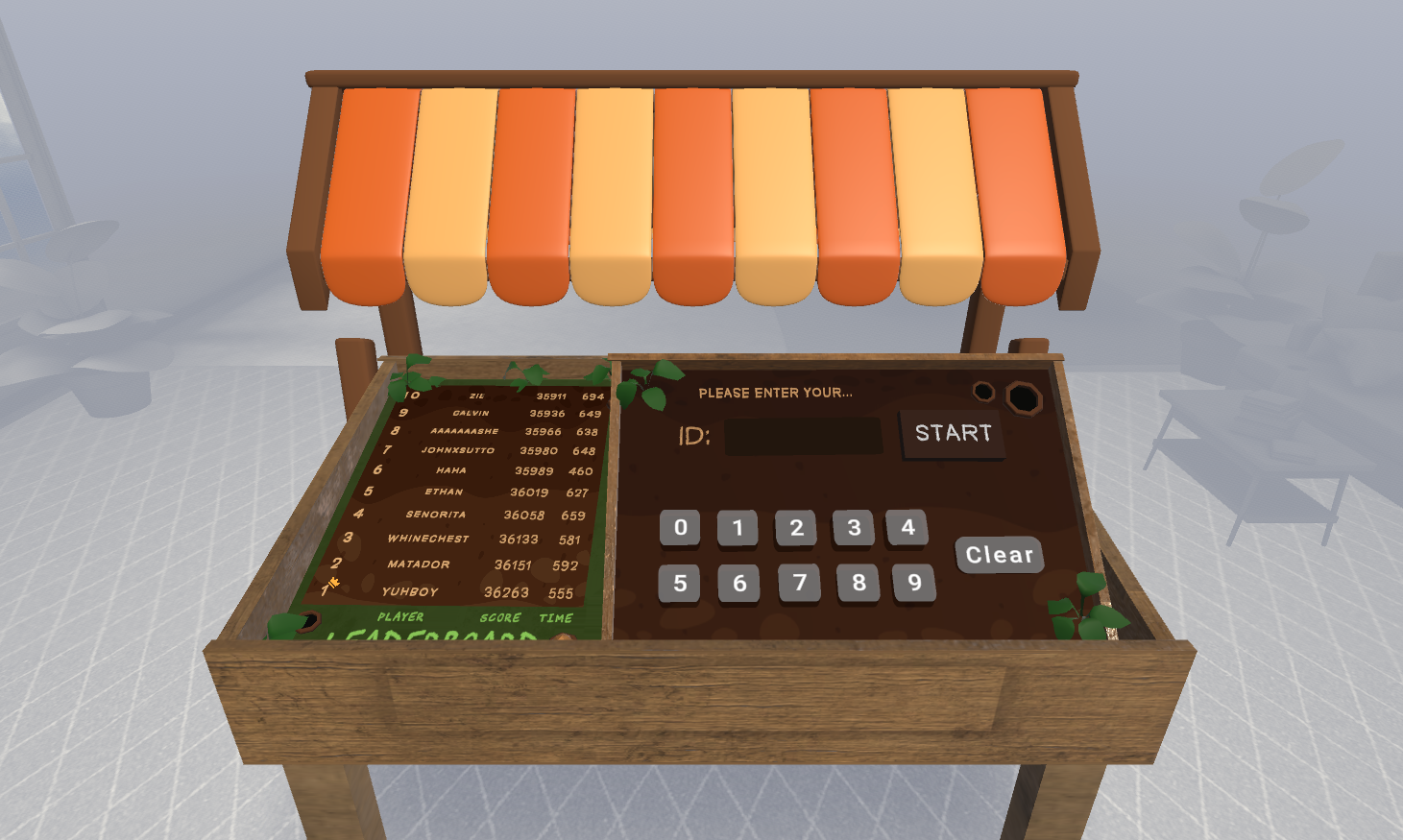}
  \caption{A 3D planting bed as the interactive VE.}
  \label{fig:planting-bed}
\end{figure}

The evaluation environment presented six static virtual rings with a black void within the inner edges to appear as a molehill. Each ring was 0.15 m × 0.07 m × 0.15 m in size, with an outer diameter of 0.15 m and an inner diameter of 0.10 m. The rings were positioned 0.15 m apart from one another, arranged in a circular and equidistant pattern to create a symmetrical, evenly spaced configuration (see Figure \ref{fig:ring-layout}). The setup was in the center of the planting bed angled -8.27° along the x-axis, enabling a high level of comfort for participants arms and necks throughout the tasks.

\begin{figure}[ht]
  \centering
  \includegraphics[width=0.8\columnwidth]{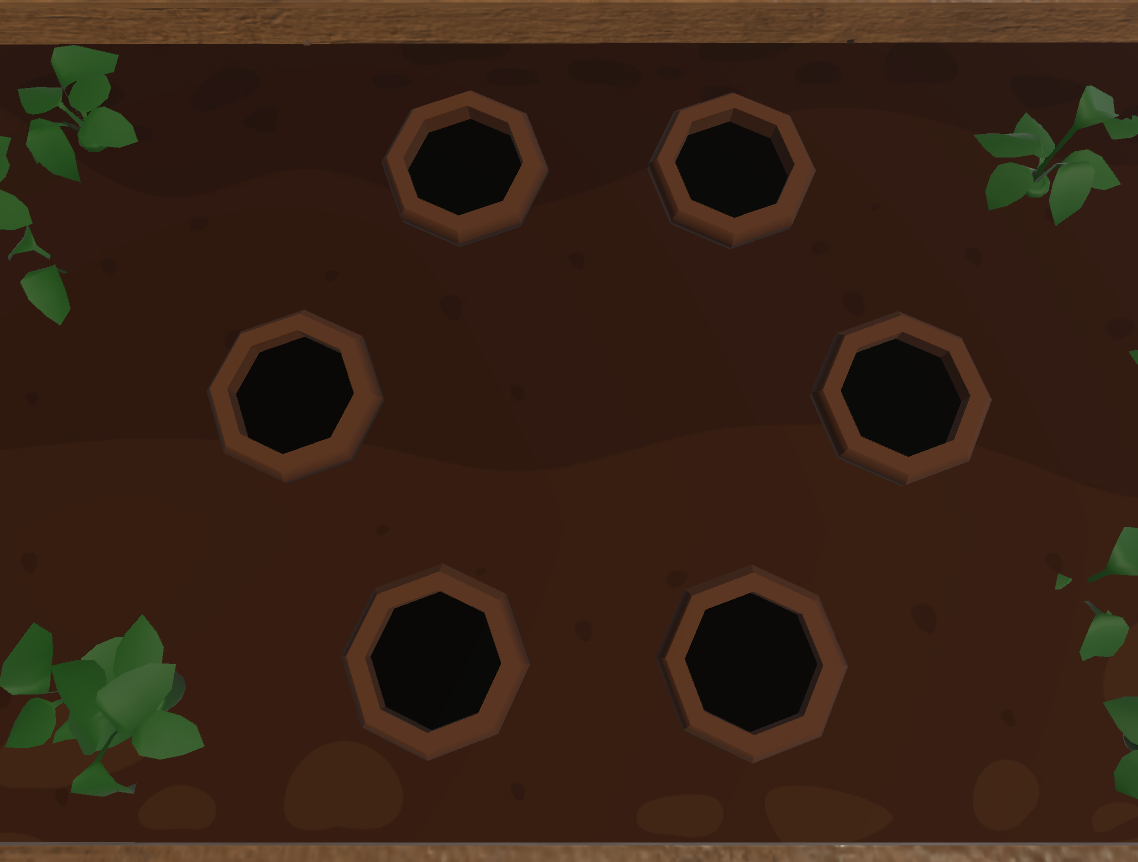}
  \caption{The six 0.15 m × 0.07 m × 0.15 m rings are placed in a circular and equidistant arrangement for the task.}
  \label{fig:ring-layout}
\end{figure}

We varied the distance to and size of the targets, in accordance with the ISO 9241-411 standard reciprocal selection task \cite{noauthor_isots_2012}. According to Fitts’ law, varying size and distance yield different task difficulty levels \cite{clark_extending_2020, kourtesis_action-specific_2022, monteiro_exploring_2023}, enabling a more accurate representation of everyday tasks participants may encounter, enhancing, enhancing ecological validity. We also incorporated other multi-directional study methods such as Shi et al. \cite{shi_expanding_2023}. We selected a target distance of 0.25 m and two diameters, 0.15 m (big) and 0.07 m (small). These yielded Fitts’ Index of Difficulty (IDs) between approximately 1.4 and 2.1. We intentionally aimed for a low-to-moderate level of difficulty on the ISO standard’s recommended range \cite{mackenzie_fitts_1992} to ensure performance quality and reduce potential confounding effects, as higher difficulty levels are associated with increased error rates and physical strain \cite{shi_expanding_2023}.

\begin{figure}[ht]
  \centering
  \begin{subfigure}[]{0.24\columnwidth}
    \includegraphics[width=\linewidth]{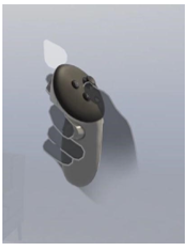}
    \caption{Hand ON and Controller ON}
    \label{subfiga:avatar-combos}
  \end{subfigure}
  \begin{subfigure}[]{0.24\columnwidth}
    \includegraphics[width=\linewidth]{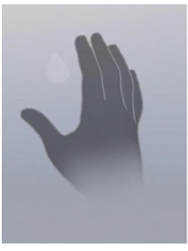}
    \caption{Hand ON and Controller OFF}
    \label{subfigb:avatar-combos}
  \end{subfigure}
  \begin{subfigure}[]{0.24\columnwidth}
    \includegraphics[width=\linewidth]{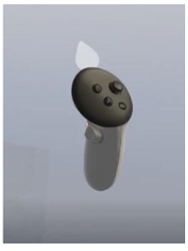}
    \caption{Hand OFF and Controller ON}
    \label{subfigc:avatar-combos}
  \end{subfigure}
  \begin{subfigure}[]{0.24\columnwidth}
    \includegraphics[width=\linewidth]{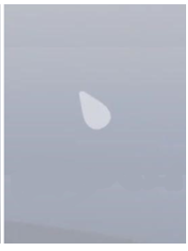}
    \caption{Hand OFF and Controller OFF}
    \label{subfigd:avatar-combos}
  \end{subfigure}
  \caption{Visual representations based on combinations of hand and controller avatar.}
  \label{fig:avatar-combos}
\end{figure}

The software also presented a different visual representation of the hand/controller avatar depending on the condition. The visual representation consisted of all possible combinations of having both the hand and controller avatars being toggled on or off. Thus, the visual representation in a condition could have both hand and controller avatars (Figure \ref{subfiga:avatar-combos}), only a hand avatar (Figure \ref{subfigb:avatar-combos}), only a controller avatar (Figure \ref{subfigc:avatar-combos}), or neither (Figure \ref{subfigd:avatar-combos}).
Regardless of visual representation, the software always displayed a small cone-shaped cursor (see Figure \ref{subfigd:avatar-combos}) to indicate the user’s hand/controller position. The software also presented the tasks’ targets and interactable objects (a hammer and a ball, used in the two different tasks) as two different sizes, big and small (see Figure \ref{fig:interactables}). The size depended on the condition.

\begin{figure}[ht]
  \centering
  \begin{subfigure}[b]{0.24\columnwidth}
    \includegraphics[height=2.5cm, width=\linewidth]{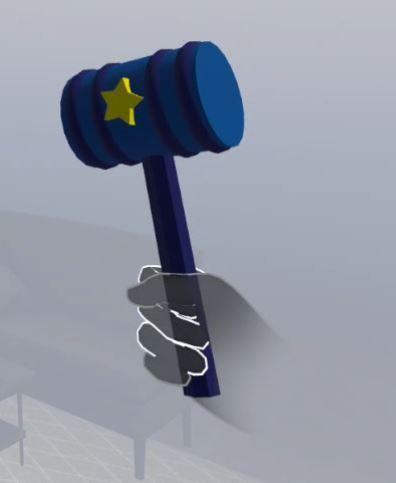}
    \caption{Hammer (Big)}
    \label{fig:interactables-a}
  \end{subfigure}
  \begin{subfigure}[b]{0.24\columnwidth}
    \includegraphics[height=2.5cm, width=\linewidth]{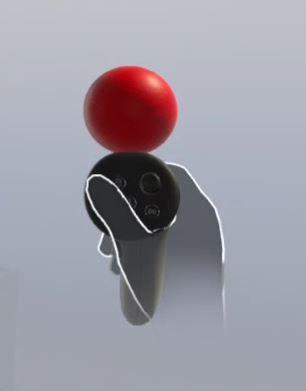}
    \caption{Ball (Big)}
    \label{fig:interactables-b}
  \end{subfigure}
  \begin{subfigure}[b]{0.24\columnwidth}
    \includegraphics[height=2.5cm, width=\linewidth]{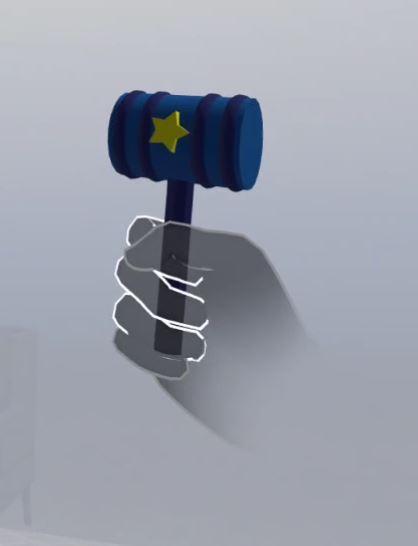}
    \caption{Hammer (Small)}
    \label{fig:interactables-c}
  \end{subfigure}
  \begin{subfigure}[b]{0.24\columnwidth}
    \includegraphics[height=2.5cm, width=\linewidth]{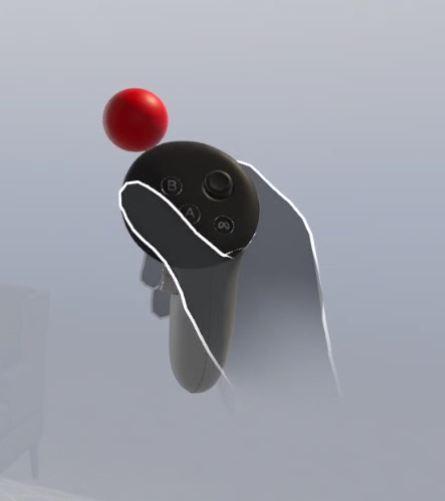}
    \caption{Ball (Small)}
    \label{fig:interactables-d}
  \end{subfigure}
  \caption{The two interactable objects used in the tasks, a hammer and a ball, shown for the big (a, b) and small (c, d) conditions.}
  \label{fig:interactables}
\end{figure}

To enhance participant engagement, we added a leaderboard to foster a sense of competition \cite{smiderle_impact_2020}. Competition-based elements, such as leaderboards (Figure \ref{fig:planting-bed}), help motivate participants by challenging their skills, encouraging peak performance \cite{halim_effectiveness_2024}.

\subsection{Procedure}

Prior to the main experiment, we conducted a pilot study with 8 participants (distinct from those in the main study) to evaluate the experimental flow and refine the task design. Each participant completed 18 trials of a \textbf{pinching} task using a single target size in a simplified virtual environment featuring floating rings. The pilot revealed that participants experienced significant difficulty and discomfort with the pinch gesture, which required high levels of precision and imposed considerable mental and motor effort. Based on these observations, we decided to include a more commonly used —\textbf{grasping}—to broaden the scope of our results. Given the challenges observed in the pinch task, we chose to have participants perform the grasp task first in the main study, allowing them to gain familiarity with the environment and reduce potential frustration before attempting the more demanding pinch task.

Upon arrival for the main experiment, participants completed a demographic questionnaire. We then demonstrated the game and explained the experimental tasks (see below). They then completed an informed consent form approved by our university research ethics board. Participants then put on the Meta Quest 3 device and the experiment started. Like the pilot study, participants performed two tasks requiring them to reach towards different rings positioned in a circle. Both tasks always commenced with the participant’s hand/controller at the start point marked with an ‘X’ at the center of the rings. The second location (i.e., the end point) was the target ring highlighted green. Distance between the start and end points was always 25 cm. Participants performed two different tasks covering the most common gestures in object manipulation: the “Whack-a-Mole” task used a grasp gesture, while the “Plug-a-Hole” task used a pinch gesture.

Informed by the pilot, we introduced two different target sizes to incorporate varying difficulty levels and improve the generalizability of the results. The virtual environment was also improved: rings were now embedded into a planting bed with moles (for grasp) and holes (for pinch), and are angled slightly for better ergonomic comfort. We increased the number of trials per condition to 24 to enhance data reliability. Lastly, we refined the success criterion for the pinch task—now, the ball must fully pass through the ring to count as a successful attempt, regardless of whether it touches the ring or not. These modifications aimed to support more natural interactions and generate more meaningful performance data across all conditions.

\paragraph{Grasp Task (Whack-a-Mole)}
In this task, participants used grasping gestures to pick up a virtual hammer (see Figure~\ref{fig:interactables-a}/\ref{fig:interactables-c}). At the beginning of the task, the hammer appeared in arm’s reach, and participants grabbed it with the current input technique.

\begin{figure}[ht]
  \centering
  \includegraphics[width=0.8\columnwidth]{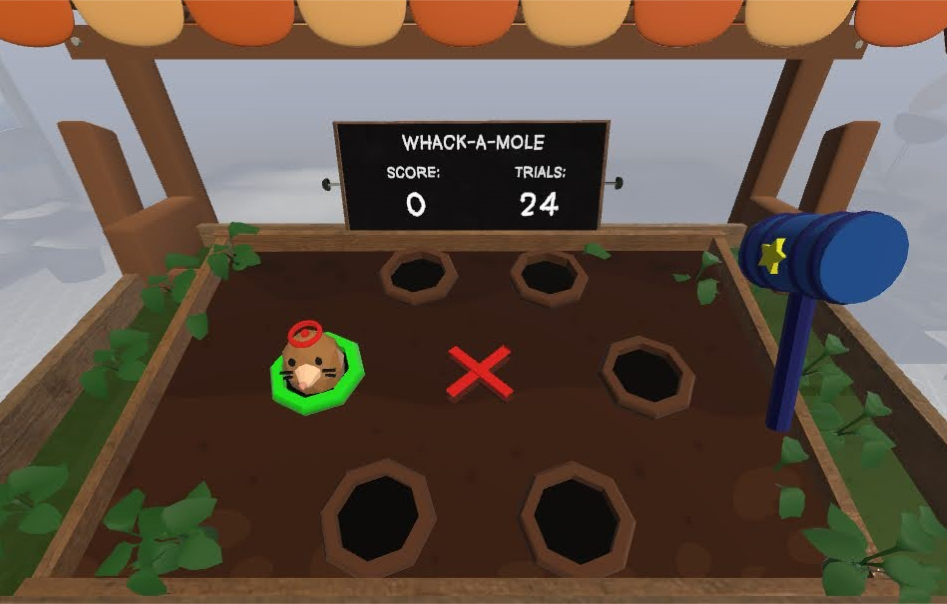}
  \caption{The Whack-a-Mole task; the blue hammer (right) and the red ‘X’ mark surrounded by the ring setup (center).}
  \label{fig:whackamole}
\end{figure}

A trial began when participants hit the ‘X’ mark at the center of the circle of rings using the hammer. This made a ring turn green, indicating it was the target. Then, a 3D model of a mole\footnote{The mole model, found within the targets, was retrieved from Sketchfab at \href{https://skfb.ly/ovrVu }{https://skfb.ly/ovrVu }.} with a visual indicator appeared at the target location. Participants moved the hammer to the target ring and struck the mole with the hammer to complete the trial. Once the hammer collided with the target mole, the system recorded the distance between the centers of the hammer’s face and mole’s head. Target ring order followed the standardized Fitts’ law reciprocal selection sequence (i.e., always across the circle of targets), starting with the top-right ring.

\paragraph{Pinch Task (Plug-a-Hole)}
In this scenario, participants used a pinch gesture to pick up and move balls. At the start, a red ball appeared in the center of the rings (i.e., at the start point) and one of the rings turned green, accompanied by a visual indicator, marking the ending point. See Figure~\ref{fig:plugahole}. Target rings were assigned in the same manner as the Whack-a-Mole task, with the fixed starting point being the left-centered ring. Picking up the ball using the pinch gesture initiated the trial. Participants then moved the ball to the target ring to complete the trial. The trial ended when the center of the ball collided with the target ring, recording the distance between the centers of the ball and ring.

\begin{figure}[ht]
  \centering
  \includegraphics[width=0.8\columnwidth]{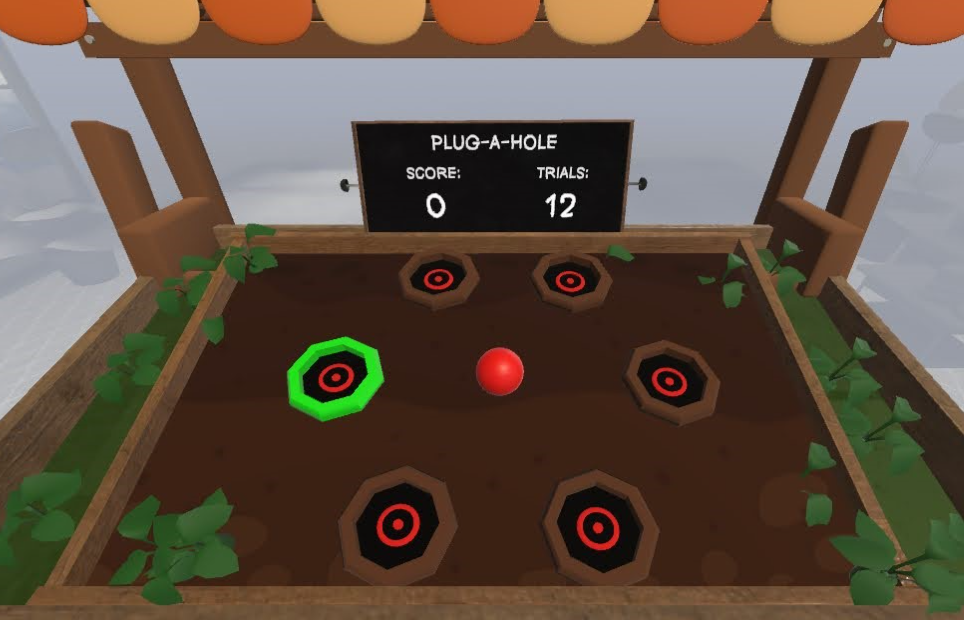}
  \caption{The Plug-a-Hole task; the red ball surrounded by the ring setup (center).}
  \label{fig:plugahole}
\end{figure}

With each new condition, participants first completed 6 non-recorded practice trials, 3 for grasp and 3 for pinch tasks. Immediately following the practice trials were the recorded trials. The recorded trials consisted of 12 trials for each of the Whack-a-Mole (grasp) task, and the Plug-A-Hole (pinch) task.

In total, they completed 24 recorded trials per condition. We included a textual and visual step-by-step guide on how to perform the task and describing the current condition. After completing all trials for a given condition, the experiment would present participants with a virtual survey. Participants answered two Likert-scale questions about their experience in the recently completed condition. Participants could take breaks at this time. After completing the survey, the experiment proceeded to the next condition. Participants answered a questionnaire at the end of the experiment related to their overall device, visual representation, sizing, and task preferences. Finally, we compensated them with \$15 and thanked them for their participation.

\subsection{Design}

Participants performed the tasks under different conditions in a $2\times2\times2\times2$ within-subjects design. Independent variables and their levels included:

\begin{itemize}
  \item \textbf{Input}: Controller, Hand Tracking 
  \item \textbf{Hand Avatar}: Hand On, Hand Off 
  \item \textbf{Controller Avatar}: Controller On, Controller Off
  \item \textbf{Target Size}: Big, Small
\end{itemize}

The ordering of the 16 combinations of these four factors was counterbalanced according to a balanced Latin square. Given that they completed 24 recorded trials (12 with the grasp task, and 12 with the pinch task) for each of the 16 conditions ($2\times2\times2\times2$ design), participants completed a total of 384 trials (16 conditions $\times$ 24 trials), consisting of 192 grasping trials (Whack-a-Mole) and 192 pinching trials (Plug-a-Hole). The value of total trials excludes the practice trials. The experiment took around 40 minutes on average to complete.

The dependent variables included completion time and accuracy. Completion time (seconds) was the time taken from the start of the task (i.e., touching the red 'X' in the grasp task, or pinching the red ball in the center of the rings in the pinch task) to the end of the task (i.e., hitting the mole in the grasp task, or placing the ball in the hole in the pinch task). Accuracy (cm) was the difference between the center of the hammer’s face or ball and the center of the target upon completion of the task. The collected data of our experiment is available at \href{https://osf.io/urk43/?view\_only=19cf112eb7d4440ead996f2719747223}{Open Science Framework website}\footnote{\href{https://osf.io/urk43/?view\_only=19cf112eb7d4440ead996f2719747223}{https://osf.io/urk43/?view\_only=19cf112eb7d4440ead996f2719747223}}.

\section{Results}

We conducted a four-way within-subjects repeated measures ANOVA to analyze completion time and accuracy for both the grasping (Whack-a-Mole) and pinching (Plug-a-Hole) tasks. For all results figures (Figure~\ref{fig:grasp-completion}, Figure~\ref{fig:grasp-accuracy}, Figure~\ref{fig:pinch-completion}, and Figure~\ref{fig:pinch-accuracy}), error bars show ±1 SE and black horizontal lines (\tikz[baseline=-0.5ex]{\fill (0,0) circle (1pt);\draw (0,0) -- (0.4,0);\fill (0.4,0) circle (1pt);}) between bars depict pairwise significant differences via Bonferroni post-hoc tests ($p < .05$). Each bar represents the average performance of the merged big and small target size conditions, grouped by the same input and visual representation combination. This was also applied to Figure~\ref{fig:figure11}, Figure~\ref{fig:figure12}, and Figure~\ref{fig:figure15}. We also analyzed participants’ subjective responses using the Aligned Rank Transform (with ANOVA) \cite{wobbrock_aligned_2011} test with Bonferroni correction for post-hoc testing \cite{elkin_aligned_2021} to compare the different interactive approaches.

\subsection{Whack-a-Mole: Grasp Gesture}

\textbf{Completion Time.} There were three significant main effects with the Grasp task. First, there was a significant main effect for input on completion time ($F_{1,32}=55.15$,$p<.001$,$\eta_p^2=.63$), with a mean time of 0.78~s (SD = 0.057~s) with controller, compared to a mean time of 1~s (SD = 0.083~s) with hand tracking, as seen in Figure~\ref{fig:grasp-completion}. The main effect for controller avatar also had a significant effect on completion time ($F_{1,32}=5.29$, $p<.028$, $\eta_p^2=.14$). Controller avatar \textit{on} yielded faster completion times (mean of 0.87~s, SD = 0.124~s) versus \textit{off} (mean of 0.92~s, SD = 0.139~s). Finally, target size had a significant main effect on completion time ($F_{1,32}=19.13$,$p<.001$,$\eta_p^2=.38$). Tasks with big targets were faster (mean of 0.84~s, SD = 0.108~s) than those with small targets (mean of 0.94~s, SD = 0.133~s). The main effect for hand avatar was not significant ($F_{1,32}=.06$,$p=.81$,$\eta_p^2=.002$), suggesting its presence or absence had comparatively little impact on participants’ performance speed. No interaction effects were found.

\begin{figure}[ht]
\centering
\small
\includegraphics[width=0.95\columnwidth]{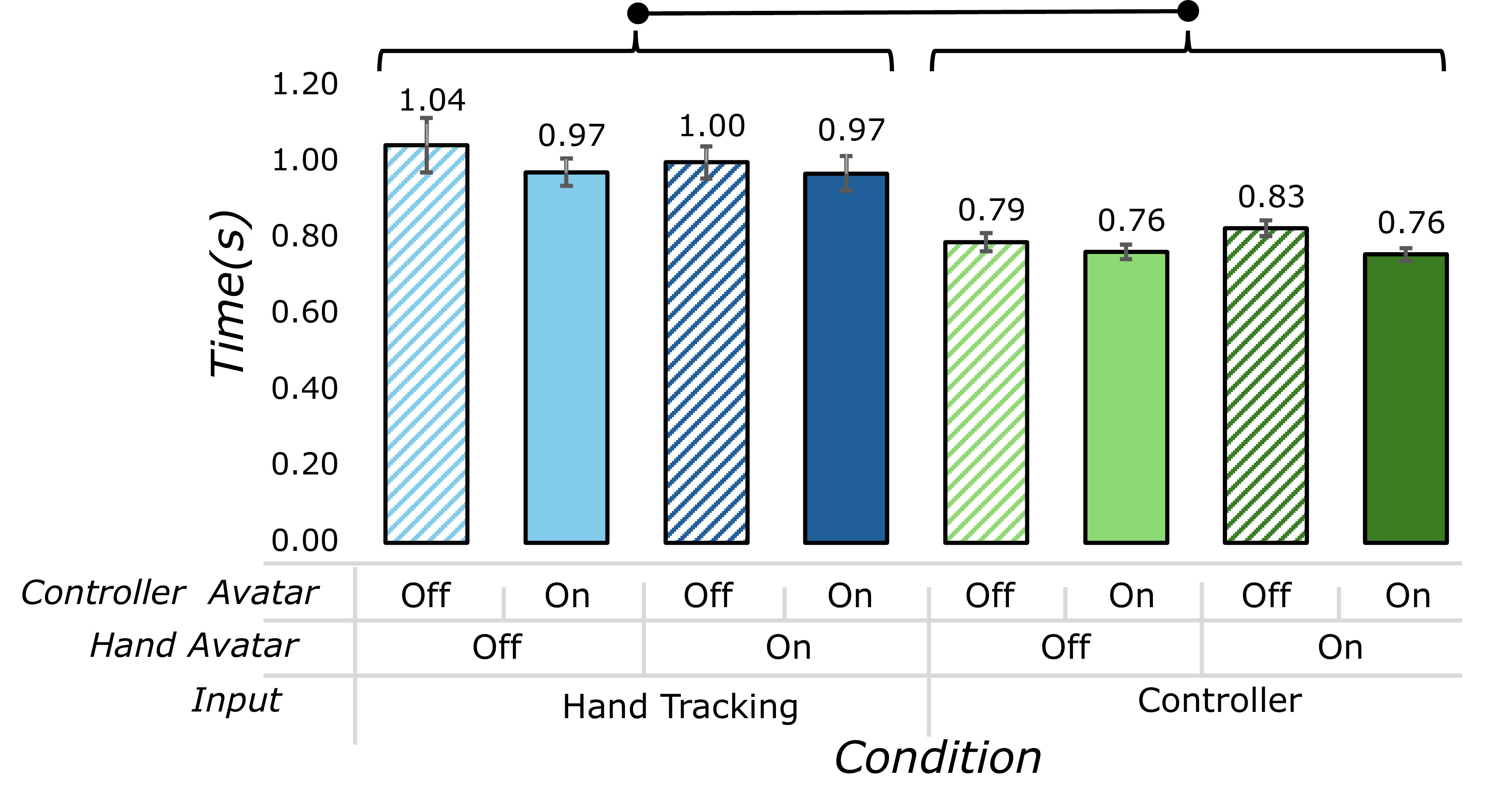}
\caption{Mean completion time by condition for the Grasp task. Only significant differences between the input is shown for clarity.}
\label{fig:grasp-completion}
\end{figure}

\textbf{Accuracy.} ANOVA revealed significant main effects for accuracy for all independent variables, including input ($F_{1,32}=21.80$,$p<.001$,$\eta_p^2=.41$), hand avatar ($F_{1,32}=58.82$,$p<.001$,$\eta_p^2=.65$), controller avatar ($F_{1,32}=6.90$,$p<.013$,$\eta_p^2=.18$), and target size ($F_{1,32}=201.24$,$p<.001$,$\eta_p^2=.86$). For input, hand tracking (mean of 5.2~cm, SD = 0.19~cm) was more accurate than controller (mean of 5.7~cm, SD = 0.19~cm). With hand avatar \textit{on}, participants were more accurate (mean distance of 5.2~cm, SD = 0.18~cm) compared to not having the hand avatar (mean of 5.6~cm, SD = 0.18~cm). Having controller avatar \textit{off} slightly outperformed having controller avatar \textit{on} (mean of 5.3~cm, SD = 0.17~cm vs. 5.5~cm, SD = 0.16~cm, respectively). Small targets had greater accuracy (mean of 4~cm, SD = 0.03~cm), than large targets (mean of 6.9~cm, SD = 0.07~cm).

There was a significant interaction effect for \textit{hand avatar} $\times$ controller avatar ($F_{1,32}=26.78$,$p<.001$,$\eta_p^2=.46$), see Figure~\ref{fig:grasp-accuracy}. Pairwise comparisons revealed differences across all combinations except the combinations of controller avatar with and without hand avatar. Visualization with only hand avatar was the most accurate (mean = 5.5~cm, SD = 0.21~cm); having then both controller and hand avatars \textit{on} was next most accurate (mean = 5.5~cm, SD = 0.21~cm). Having both avatars \textit{off} the was least accurate (mean = 5.8~cm, SD = 0.19~cm).

More importantly, there was a three-way interaction effect between \textit{input} $\times$ \textit{hand avatar} $\times$ \textit{controller avatar} ($F_{1,32}=10.52$,$p<.003$,$\eta_p^2=.25$), as seen in Figure~\ref{fig:grasp-accuracy}. When not using \textit{any} visual avatar representation (i.e., both controller avatar and hand avatar \textit{off}), hand tracking (mean = 5.4~cm, SD = 0.2~cm) offered better accuracy than the controller (mean = 6.2~cm, SD = 0.2~cm). Similarly, when both hand and controller avatars were \textit{on}, hand tracking (mean = 5.2~cm, SD = 0.2~cm) offered better accuracy than the controller (mean = 5.9~cm, SD = 0.2~cm).

\begin{figure}[ht]
  \centering
  \small
    \includegraphics[width=0.95\columnwidth]{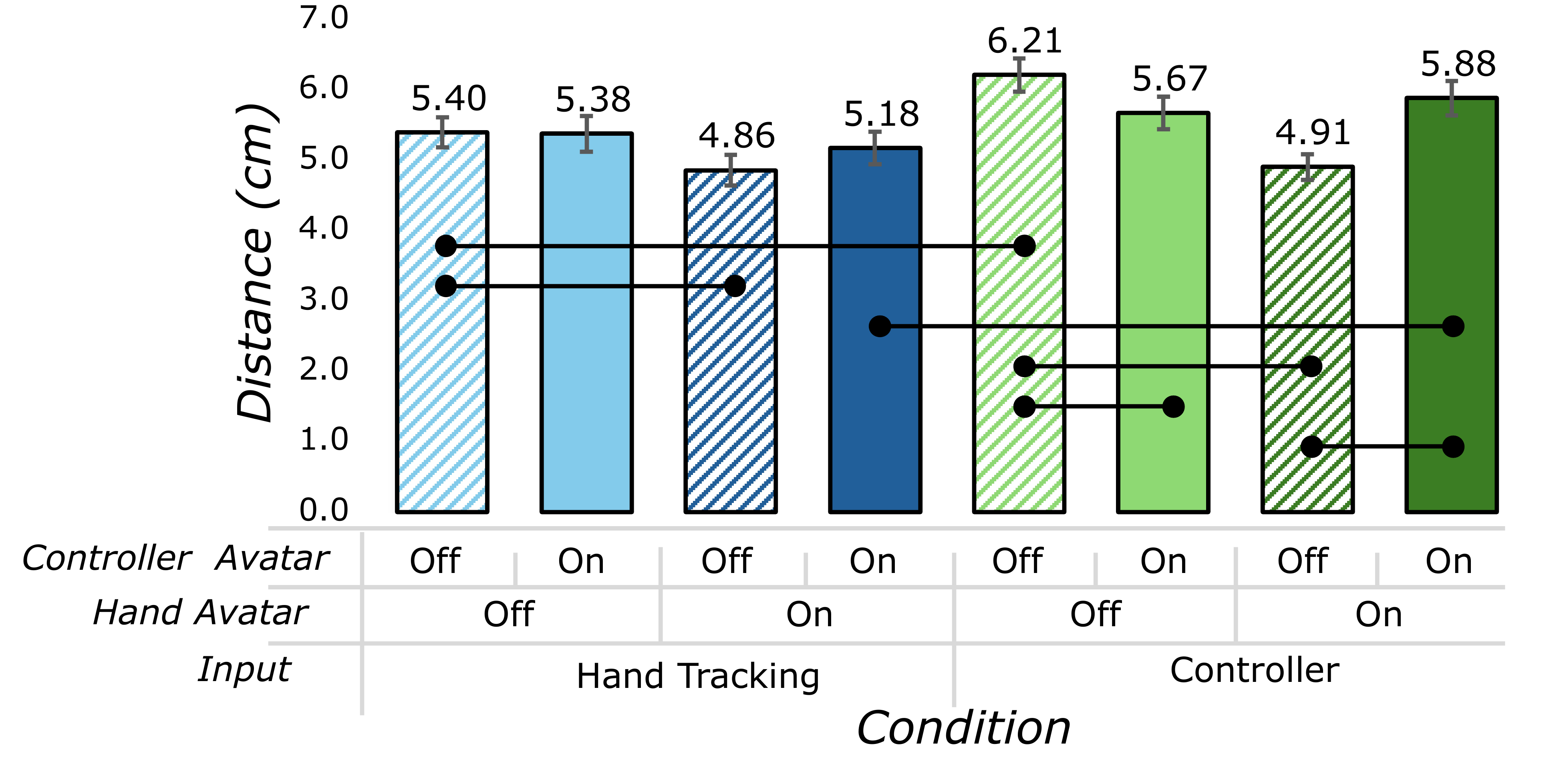}    
  \caption{Mean accuracy by condition for the Grasp task.}
  \label{fig:grasp-accuracy}
\end{figure}

Visual representations played a key role with controller input where using no visual representation (i.e., both controller and hand avatars \textit{off}) yielded worse accuracy (mean = 6.2~cm, SD = 0.2~cm) than having only the hand avatar \textit{on} (mean = 4.9~cm, SD = 0.16~cm) or only the controller avatar \textit{on} (mean = 5.7~cm, SD = 0.2~cm). To our surprise, having both avatars (mean = 5.9~cm, SD = 0.2~cm) seemed to have a negative effect compared to having only the hand avatar (mean = 4.9~cm, SD = 0.16~cm).

\subsection{Plug-a-Hole: Pinch Gesture}

\textbf{Completion time.} For the Pinch task trials (Figure~\ref{fig:pinch-completion}), ANOVA revealed relatively few significant effects compared to the Grasp trials. There were significant main effects for input ($F_{1,32}=79.39$,$p<.001$,$\eta_p^2=.71$) and size ($F_{1,32}=5.62$,$p<.024$,$\eta_p^2=.15$). Controller input was faster (mean = 0.91~s, SD = 0.037~s) than hand tracking (mean = 1.27~s, SD = 0.05~s). Moreover, big targets yielded a faster time (mean = 1.05~s, SD = 0.047~s) than small targets (mean = 1.14~s, SD = 0.039~s). No interaction effects were found.

\begin{figure}[ht]
  \centering
  \small
    \includegraphics[width=0.95\columnwidth]{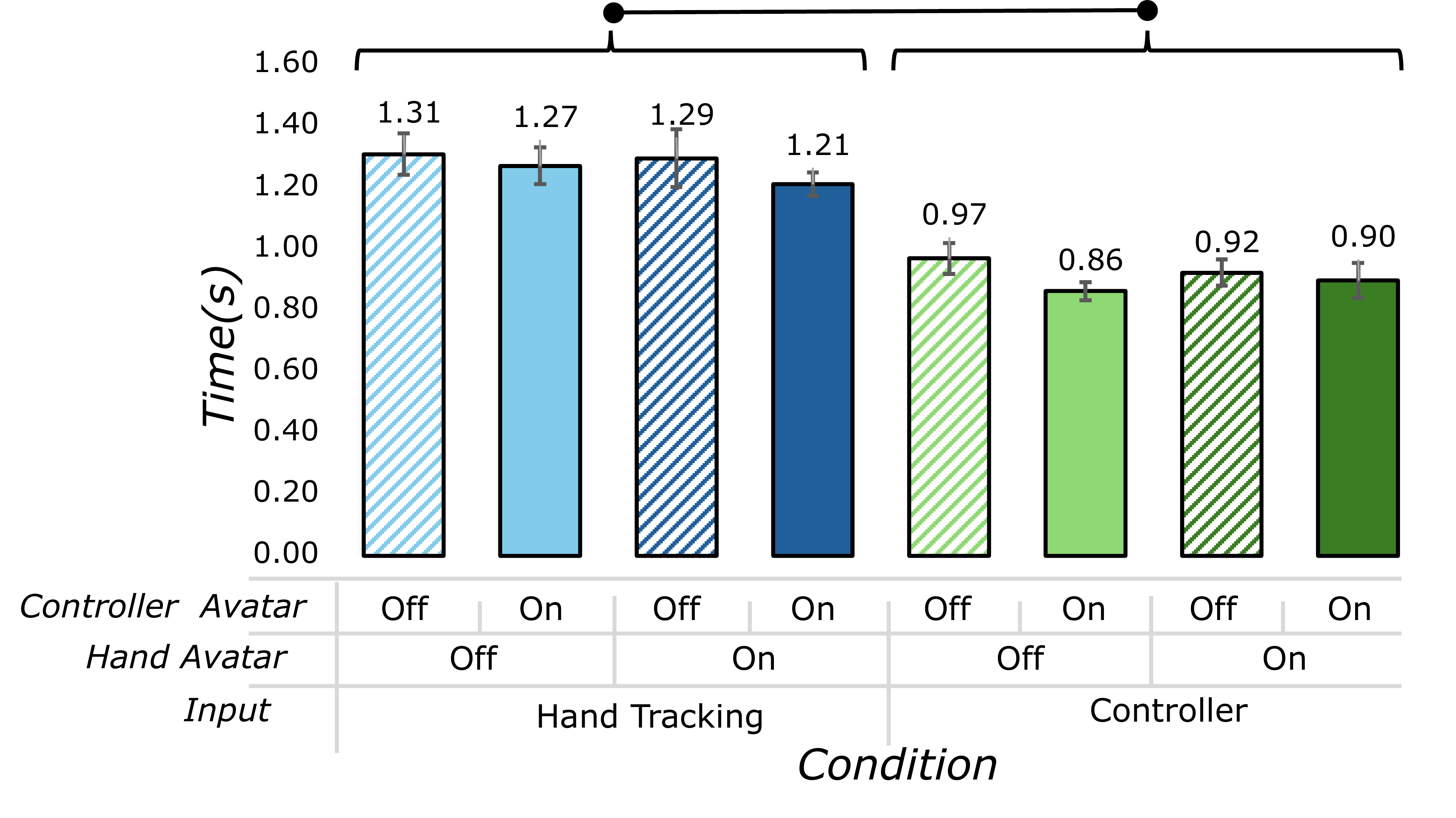}
  \caption{Mean completion time by condition for the Pinch task.}
  \label{fig:pinch-completion}
\end{figure}

\textbf{Accuracy.} Accuracy results for Pinch trials are seen in Figure~\ref{fig:pinch-accuracy}. ANOVA revealed significant main effects for input ($F_{1,32}=75.07$,$p<.001$,$\eta_p^2=.70$), controller avatar ($F_{1,32}=31.41$,$p<.001$,$\eta_p^2=.50$), and target size ($F_{1,32}=171.90$,$p<.001$,$\eta_p^2=.84$). The main effect for hand avatar was not significant ($F_{1,32}=1.55$,$p=.22$,$\eta_p^2=.05$) suggesting that it had limited impact on accuracy. Hand tracking input yielded better accuracy (mean of 3.1~cm, SD = 0.05~cm), compared to controller input (mean of 3.7~cm, SD = 0.08~cm). Accuracy was also higher when the controller avatar was \textit{on} (mean = 3.3~cm, SD = 0.06~cm) than \textit{off} (mean = 3.6~cm, SD = 0.09~cm). Similarly, smaller targets offered better accuracy (mean of 2.9~cm, SD = 0.03~cm), than larger targets (mean of 4~cm, SD = 0.06~cm).

\begin{figure}[ht]
  \centering
\small
    \includegraphics[width=0.95\columnwidth]{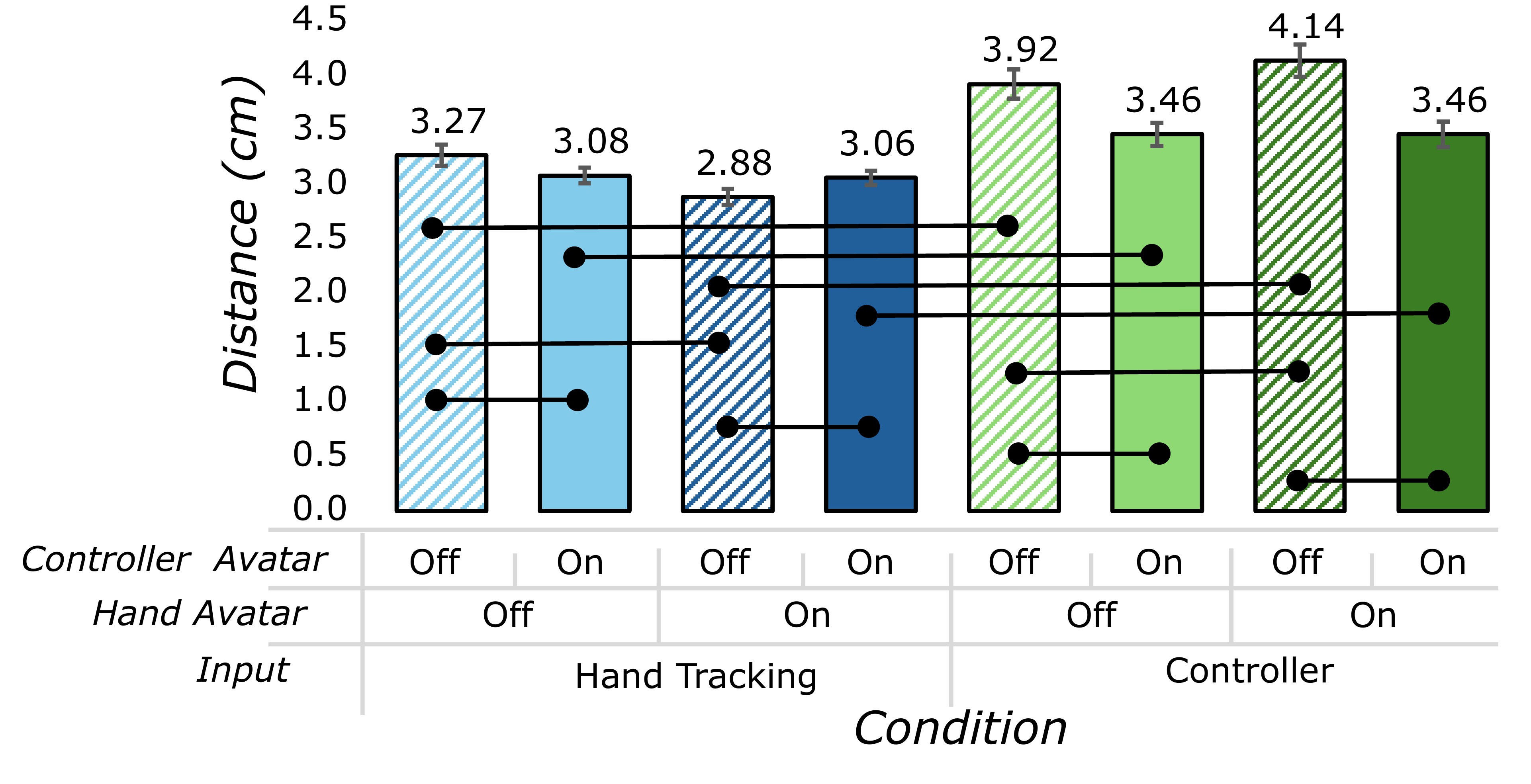}
  \caption{Mean accuracy by condition for the Pinch task.}
  \label{fig:pinch-accuracy}
\end{figure}

The \textit{Input} $\times$ \textit{Hand Avatar} interaction effect was statistically significant ($F_{1,32} = 19.46$, $p < .001$, $\eta_p^2 = .38$). Pairwise comparisons revealed differences across all combinations. The most accurate condition was hand tracking with hand avatar \textit{on} (mean = 2.97 cm, SD = 0.06 cm), followed by hand tracking with hand avatar \textit{off} (mean = 3.17 cm, SD = 0.07 cm), then controller input with hand avatar \textit{off} (mean = 3.69 cm, SD = 0.10 cm). The least accurate was controller input with a mismatched hand avatar \textit{on} (mean = 3.80 cm, SD = 0.12 cm). 
The \textit{Input} $\times$ \textit{Controller Avatar} interaction was also statistically significant ($F_{1,32} = 35.37$, $p < .001$, $\eta_p^2 = .31$). Pairwise comparisons showed significant differences across all combinations except between hand tracking and controller avatar \textit{on} and \textit{off}. In this case, hand tracking with controller avatar was slightly more accurate (mean = 3.07 cm, SD = 0.06 cm) than without (mean = 3.07 cm, SD = 0.08 cm), though the difference was not statistically significant. In contrast, controller input with controller avatar \textit{on} offered better accuracy (mean = 3.46 cm, SD = 0.10 cm) than with controller avatar \textit{off} (mean = 4.03 cm, SD = 0.13 cm).

There was a significant three-way interaction between \textit{Input} $\times$ \textit{Hand Avatar} $\times$ \textit{Controller Avatar} ($F_{1,329} = 15.84$, $p < .001$, $\eta_p^2 = .33$), as seen in Figure~\ref{fig:pinch-accuracy}. Post-hoc comparisons revealed consistent differences between input types across all hand avatar $\times$ controller avatar combinations, confirming the main effect of input: hand tracking was more accurate than controller input in every case. Differences between controller avatar conditions were also found across all input $\times$ hand avatar combinations, consistent with a main effect, although the direction varied depending on the combination. For instance, controller avatar \textit{on} improved accuracy when using hand tracking with hand avatar \textit{off} (mean = 3.08 cm, SD = 0.07 vs. mean = 3.27 cm, SD = 0.09), controller input with hand avatar \textit{off} (mean = 3.46 cm, SD = 0.12 vs. mean = 3.92 cm, SD = 0.13), and controller input with hand avatar \textit{on} (mean = 3.46 cm, SD = 0.12 vs. mean = 4.14 cm, SD = 0.14). In contrast, when using hand tracking with hand avatar \textit{on}, also having the controller avatar \textit{on} resulted in worse accuracy (mean = 3.06 cm, SD = 0.07) than having the controller avatar \textit{off} (mean = 2.88 cm, SD = 0.07). This suggests that for hand tracking, a matching representation (hand avatar only) leads to better accuracy than a mismatched one (hand + controller avatar), and a mismatched (controller avatar only) is preferable to having no avatar at all.
Conversely, for controller input, the presence of the controller avatar improves accuracy regardless of whether the hand avatar is present or not. 

Finally, we found significant differences between hand avatar conditions when the controller avatar was \textit{off}: with hand tracking, having only a hand avatar led to better accuracy (mean = 2.88 cm, SD = 0.07) than no avatar (mean = 3.27 cm, SD = 0.09), whereas with controller input, accuracy was higher without any avatar (mean = 3.92 cm, SD = 0.13) than when using only the mismatched hand avatar (mean = 4.14 cm, SD = 0.14).  This suggests that with hand tracking, a mismatched representation (only hand avatar) is better than no avatar at all. In contrast, with controller input, it is preferable to have \textit{no avatar at all} rather than a mismatched one.

\subsection{In-study Subjective Responses}

Upon completing each condition, participants answered the question \textit{“How real/natural did the \textasteriskcentered input type\textasteriskcentered{} paired with the \textasteriskcentered avatar\textasteriskcentered{} feel?”}. Participants' responses were ranked on a 5-point Likert scale where 1 meant “Felt Unfamiliar/Artificial” and 5 meant “Felt Like My Real Hand.” 
We analyzed the data using an Aligned Rank Transform (ART) ANOVA~\cite{wobbrock_aligned_2011}, revealing significant effects of Input ($F_{1,329} = 76.44$, $p < .001$), Hand Avatar ($F_{1,329} = 4.23$, $p = .040$), and Controller Avatar ($F_{1,329} = 6.61$, $p = .011$). We examined the proportion of positive responses (scores 4 and 5). Conditions using controller input were perceived as more realistic, with 77.1\% of responses being positive, compared to 62.0\% for hand tracking. With hand avatar \textit{on}, 71.9\% of responses were positive, versus 67.2\% when hand avatar was \textit{off}. Finally, for the controller avatar, the proportion of positive responses was similar whether the avatar was shown (69.8\%) or not (69.3\%). However, the proportion of strongly negative responses (scores 1 and 2) differed: 13.5\% when the controller avatar was present, compared to 10.4\% when it was absent.

The two-way interaction \textit{input} $\times$ \textit{hand avatar} ($F_{1,329} = 10.89$, $p = .001$) was found to be significant. Post-hoc comparisons were conducted using the Aligned Rank Transform Procedure for Multifactor Contrast Tests (ART-C)~\cite{elkin_aligned_2021} with Bonferroni correction. All controller input conditions differed significantly from all hand tracking conditions (all $p < .01$). Controller input showed higher perceived realism with hand avatar \textit{on} (78.1\% positive responses) and \textit{off} (76.0\%), compared to hand tracking, with hand avatar \textit{on} (65.6\%) and \textit{off} (58.3\%). Similarly, results showed the interaction Input $\times$ Controller Avatar ($F_{1,329} = 5.49$, $p = .020$) to be significant. Post-hoc test showed that both controller input conditions differed significantly from both hand tracking conditions (all $p < .001$). Controller input resulted in more positive responses both when the controller avatar was \textit{on} (79.2\%) or \textit{off} (75.0\%), compared to hand tracking, with controller avatar \textit{on} (60.4\%) and \textit{off} (63.5\%).

\begin{figure}[ht]
\centering
\small
\includegraphics[width=\columnwidth]{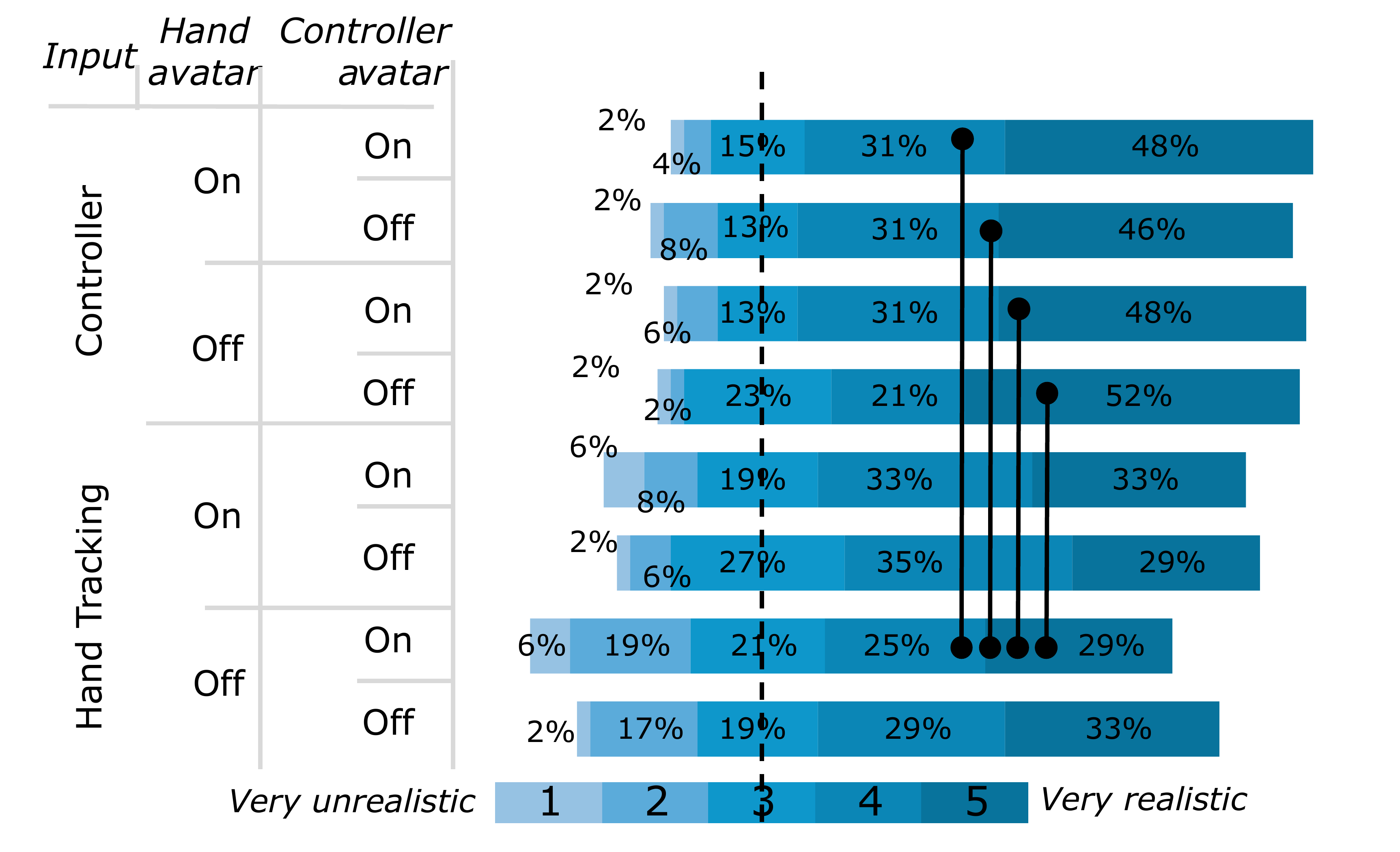} %
\caption{Participants' response on perceived sense of realism for each condition. Black vertical bars show pairwise differences via post-hoc tests ($p < .05$).}
\label{fig:figure11}
\end{figure}

A significant three-way interaction was observed for \textit{input} $\times$ \textit{hand avatar} $\times$ \textit{controller avatar} ($F_{1,329} = 7.14$, $p = .008$). Post-hoc comparisons showed significant differences between all controller combinations and one hand tracking condition ($p < .01$). Controller input had higher perceived realism with both avatars \textit{on} (79.2\% positive), with only the hand avatar \textit{on} (77.1\%), with only the controller avatar \textit{on} (79.2\%), and with both avatars \textit{off} (72.9\%) compared to hand tracking using only the mismatched controller avatar (54.2\%), which was perceived as the least realistic condition overall (see Figure~\ref{fig:figure11}).

\begin{figure}[ht]
\centering
\small
\includegraphics[width=\columnwidth]{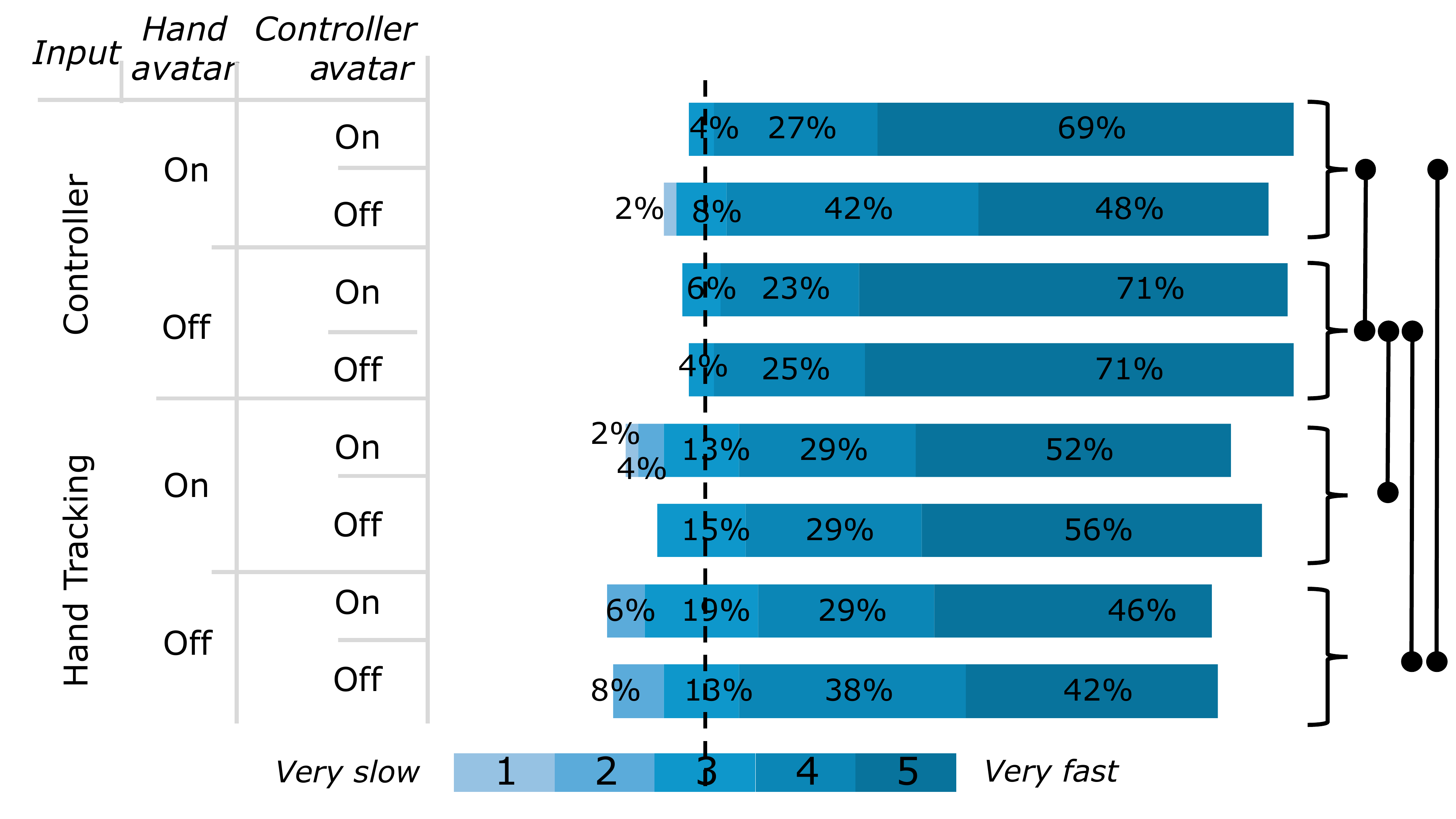} %
\caption{Participants' response on perceived adaptability for each condition. Black vertical bars show pairwise differences between non-different groups designated by parentheses via post-hoc tests ($p < .05$).}
\label{fig:figure12}
\end{figure}

Participants also answered \textit{“How quickly did you get used to \textasteriskcentered input type\textasteriskcentered{} paired with the \textasteriskcentered avatar\textasteriskcentered{}?”} on a 5-point Likert scale, where 1 meant “Took Me A While” and 5 meant “Immediately.” An ART ANOVA revealed significant main effects of Input ($F_{1,329} = 12.00$, $p < .001$) and Controller Avatar ($F_{1,329} = 4.65$, $p = .032$). For the Input, controller input led to faster adaptation overall, with 93.8\% of responses indicating high familiarity (scores 4–5) compared to hand tracking with 80.2\% positive scores and 5.2\% indicating perceived slow adaptation. For the Controller Avatar, when it was \textit{off}, 87.5\% of responses indicated fast adaptation compared to 86.5\% when it was \textit{on}; in contrast, low adaptation (scores 1–2) was reported in 2.6\% of cases without the avatar and 3.1\% with it.

A significant two-way interaction was also found for \textit{input} $\times$ \textit{hand avatar} ($F_{1,329} = 10.46$, $p = .001$). Post-hoc comparisons (ART-C with Bonferroni correction) revealed significant differences between controller input without hand avatar and all other conditions (all $p < .05$), and between controller with hand avatar and hand tracking without hand avatar ($p = .028$). Controller input without hand avatar had the highest rate of fast adaptation, with 94.8\% of responses in the 4–5 range, followed by controller input with hand avatar (92.7\%), hand tracking with hand avatar (83.3\%), and hand tracking without hand avatar (77.1\% fast), as seen in Figure~\ref{fig:figure12}.

\subsection{Post-Study Survey}

At the end of the experiment, participants completed a survey about their experience with the input devices and overall preferences.

\textbf{Input devices.} Participants rated how successful they felt using the input devices on a 5-point Likert scale, where 1 meant “Not Successful” and 5 meant “Very Successful.” See Figure~\ref{fig:figure13}. A Wilcoxon signed-rank test ($Z = -4.57$, $p < .001$) revealed that participants perceived themselves as significantly more successful with controller input, with 92\% of them rating between 4 and 5 compared to the hand tracking, where these scores obtained 69\%.

\begin{figure}[ht]
\centering
\small
\includegraphics[width=\columnwidth]{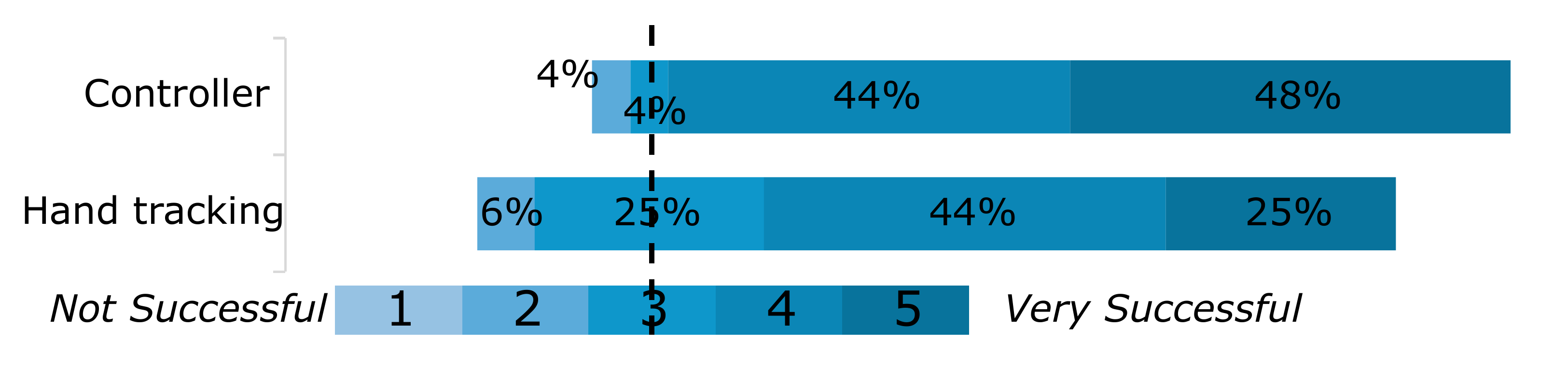} 
\caption{Participants’ answers related to the perceived self-performance of the tasks on the different input devices.}
\label{fig:figure13}
\end{figure}

Participants also rated the perceived difficulty level when performing the tasks with each input type. See Figure~\ref{fig:figure14}. A Wilcoxon signed-rank test ($Z = -4.61$, $p < .001$) revealed that participants perceived tasks as less difficult when using the controller input, with 88\% of responses between 1 and 2, compared to 40\% for the hand tracking results.

\begin{figure}[ht]
\centering
\small
\includegraphics[width=0.95\columnwidth]{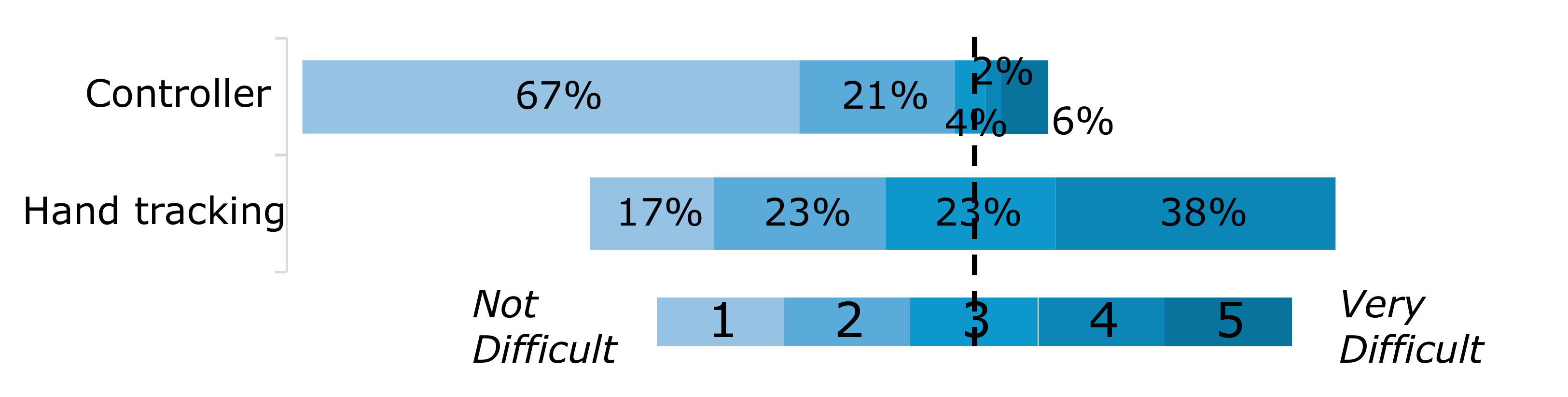} %
\caption{Participants’ answers related to the perceived tasks’ difficulty on the different input devices.}
\label{fig:figure14}
\end{figure}

Participants indicated their most and least preferred avatar representations for each input type. See Figure~\ref{fig:figure15}. With controller input, half of the participants preferred having both hand and controller avatars, while the least preferred option was having no avatar at all. For hand tracking, 44\% (21 participants) selected having both avatars as preferred option, followed closely by 40\% (19 participants) who preferred only the hand avatar. Similar to controllers, the least preferred option was having no avatar, with 42\% (20 participants) expressing dissatisfaction with this condition.

\begin{figure}[ht]
\centering
\small
\includegraphics[width=\columnwidth]{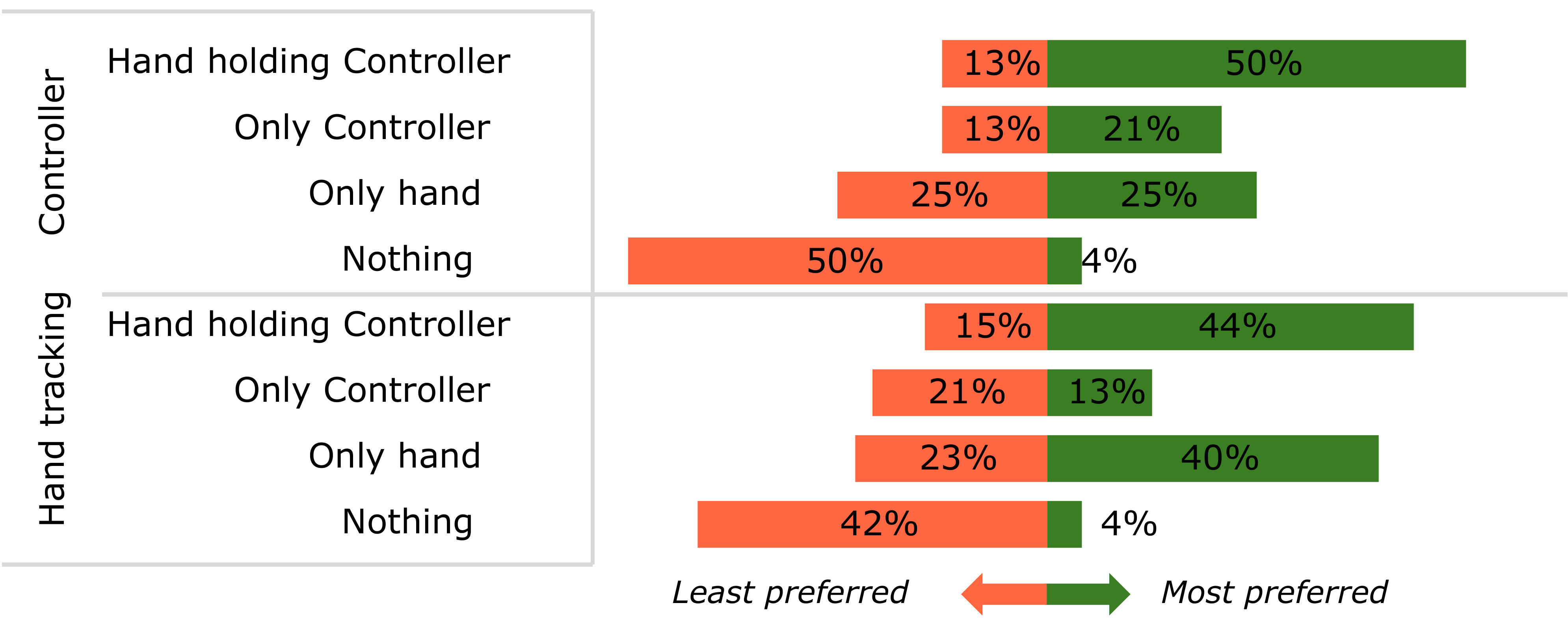} %
\caption{Participants’ most and least preferred avatar representation on each input device.}
\label{fig:figure15}
\end{figure}

This preference distribution was reflected in participants’ open-ended feedback. For controller input, many found the paired hand and controller avatar natural and intuitive. P27 and P7 commented, \textit{“I liked how my hand and controller were represented so it felt very similar to reality… when what I was holding wasn’t there it felt like something was missing,”} and\textit{ “It felt natural and it looked easy to control and aim accurately,”} respectively. Furthermore, several participants associated the pairing with \textit{“more control”} (P7, P15, P18, P28, P35), \textit{“comfort”} (P10, P22, P23, P33), and \textit{“ease of use”} (P8, P26, P30, P46, P48). While a few participants preferred the controller avatar alone, they noted familiarity and simplicity. For example, P20 noted, \textit{“I felt really natural holding a controller and seeing a controller as my avatar. It just felt right.”} In contrast, the no-avatar approach was often described as disorienting. P9 reported, \textit{“I didn’t feel like I was using anything… it’s like I didn’t have much coordination,”} while P7 commented, \textit{“It was hard to keep track of it and it was also difficult to troubleshoot and find out what the problem was.”} The hand avatar alone received mixed responses. Some participants appreciated the familiar and realistic visual as it was \textit{“more convenient”} (P41), while others felt it \textit{“made it feel much more like a video game”} (P12) than reality.

A similar pattern emerged with hand tracking. Participants preferred the visual representation of their hands, either alone or holding a controller avatar. P24 explained, \textit{“It just felt the most natural and it mimicked my real hands,”} while P7 said, \textit{“I preferred when it was just my hand, as the motions mirrored my hand’s motions identically.”} The no-avatar approach was again the most criticized, described as \textit{“confusing”} (P1, P9, P37), \textit{“no sense of navigation”} (P3, P20, P33, P41, P46), and \textit{“less immersive”} (P12, P19, P34). For instance, P27 commented, \textit{“I felt like I was just grabbing nothing.”} The controller avatar also received participant dissatisfaction. P21, P6, and P14 questioned its relevance: \textit{“Why would I want a controller visual if I’m using my hand?”}, \textit{“It felt pointless and dumb,”} and \textit{“My hand felt very useless,”} respectively, highlighting the disruption of incongruent visuals on usability.

\textbf{Participant Preferences.} The questionnaire included questions related to preferences regarding certain aspects of the experience. Some notable points include: 35 participants (73\%) found the larger target made the task easier, while 13 (27\%) preferred smaller targets. A similar trend was observed in task speed, with 39 participants (81\%) reporting they were able to complete the task faster with larger targets, whereas 9 (19\%) indicated the opposite. When asked about their preferred combination of input device and visual representation, 35 participants (73\%) favored using controllers. Among them, 16 preferred both avatars, 8 only the hand avatar, 7 only the controller avatar, and 4 preferred no avatars. Meanwhile, 13 participants (27\%) preferred hand tracking, with 5 favoring only the hand avatar, 4 selecting both avatars, 3 preferring only the controller avatar, and 1 opting for no avatar at all. Lastly, the majority of participants (90\%) reported enjoying the grasping (Whack-a-Mole) task more than the pinching (Plug-a-Hole) task.

\section{Discussion}
\subsection{User Performance}

Grasping tasks were generally faster but less accurate than pinching tasks, although we did not conduct a direct statistical comparison between them. It is possible that the more demanding nature of the pinching task required greater precision, forcing participants to take more time and thereby improving accuracy.

Controller input consistently led to faster task completion than hand tracking in both grasping and pinching tasks. This is consistent with previous findings that physical controllers provide greater stability and motor control due to their tangible structure~\cite{hamad_how_2022, kilteni_sense_2012, mackenzie_fitts_1992, johnson_exploring_2023}. The physical buttons and triggers on controllers allow for immediate, discrete input; in contrast, hand tracking relies on continuous motion detection and imprecise gesture recognition~\cite{argelaguet_role_2016}.

Our findings challenge the assumption that controllers are universally better than hand tracking in terms of both speed and accuracy~\cite{hanashima_how_2022}. While controllers offered faster performance overall, they resulted in significantly lower accuracy, particularly in the pinch task, compared to hand tracking. Participants felt a greater sense of confidence and success when using controllers, reducing concerns about potential errors. When performance is timed and scored, this confidence has led participants to prioritize focus towards speed while heavily trusting the accuracy behind their actions, leading to lower task accuracy. In contrast, hand tracking did not offer such confidence, requiring more intuitive gestures that demanded finer hand adjustments~\cite{johnson_exploring_2023}. This complexity may lead to a greater sense of difficulty, increasing the fear of errors. Thus, participants prioritized minimizing mistakes by carefully monitoring and slowing their actions, sacrificing speed.

Avatar representation had a notable effect on input accuracy. In general, matching avatar representations yielded the highest accuracy, outperforming both mismatched and absent avatars. For hand tracking, even a mismatched controller avatar was better than no avatar at all. Interestingly, this pattern differed with controller input and varied depending on the task. With controller input in the grasping task, the best accuracy was achieved with only the hand avatar; in the pinching task, the same condition yielded the worst accuracy. In that case, showing only the controller avatar or both avatars offered better accuracy. These accuracy differences were not reflected in completion time, as timing trends were relatively stable across conditions. This suggests that avatar representation plays a crucial role in controller-based interactions, particularly when there is either a strong match or a noticeable mismatch between the visual and physical representation~\cite{venkatakrishnan_how_2023}.

One possible explanation lies in the role of haptic feedback and visual alignment with physical actions. In hand tracking, the absence of haptic cues forces users to rely more heavily on visual information—even if mismatched—than on alignment between visual representation and physical sensation. In contrast, controllers offer a tangible object and implicit haptic feedback. When paired with a hand avatar, the visual representation may still feel coherent because it reflects the user’s actual hand. However, this may pose limitations depending on the task. Grasping involves a gesture that naturally aligns with holding a controller, making the hand avatar appropriate in that context. In contrast, pinching requires fine motor control of the thumb and index finger, a gesture that differs significantly from gripping a controller. This likely increases the dissonance between visual and physical input when mismatches occur. Our findings partially support Venkatakrishnan et al.~\cite{venkatakrishnan_how_2023}, who noted that controller-based hand placement can negatively affect performance. We observed that this influence may be task-dependent, potentially making the performance better or worse. This variation may be explained by how well the physical gesture aligns with the virtual representation. In other words, hand tracking offers more flexibility in gesture design, but its lower reliability increases the need for visual feedback, even if mismatched. Controllers provide more robust input but only benefit from a one-handed avatar representation when the virtual gesture aligns with the physical posture of holding the controller. Otherwise, the mismatch may be counterproductive.
Finally, we note that target size had a significant impact on both completion time and accuracy. This is consistent with Fitts’ law, where bigger targets lower the target acquisition difficulty and thereby completion time. Small targets yielded higher accuracy than big targets, probably because small targets require greater precision for fine motor control movements~\cite{buckingham_hand_2021}, and the measurements in bigger targets allowed users to aim farther from the center’s target, allowing greater distances.

Our findings suggest that while controllers generally enable faster task performance, this speed can come at the cost of reduced accuracy compared to hand tracking. Avatar representation had a significant impact on accuracy, although no clear effects were observed on completion time. Hand tracking offered greater flexibility in gesture expressiveness, making it a viable option when virtual actions differ from the physical gesture of holding a controller. In contrast, controllers provide robust and fast input, but benefit from careful avatar design. Using only a hand avatar can enhance accuracy when the virtual gesture closely matches the physical interaction; otherwise, it may lead to performance degradation. In such cases, including a controller avatar—or a similarly shaped visual cue—can help align user expectations with physical constraints, potentially improving performance and user experience.

\subsection{User Experience}

The subjective results revealed a general preference for controllers over hand tracking. Mismatched avatar-input pairings negatively affected perceived realism, as evidenced by the low ratings for hand tracking paired with only the controller avatar. Participants reported similar levels of perceived adaptation speed across conditions; however, the presence of the controller avatar had little impact. Controllers were consistently rated as the easiest to adapt to, particularly when used without the hand avatar. This suggests the pairing between input type and avatar representation influences user perception of embodiment and usability, consistent with prior studies~\cite{johnson_exploring_2023, lin_need_2016, ponton_stretch_2024, venkatakrishnan_how_2023}.

Participants also expressed a preference towards the grasp tasks over the pinch tasks. We were surprised by how drastic this preference difference was. With hand tracking, this difference may be due to the grasp gesture delivering tactile feedback upon a larger area of the palm and having a lower motor demand as finger extension was not required. In contrast, the pinch gesture requires a precise interaction point between two fingers; grasping uses all fingers in a comparatively less precision-oriented gesture. As for controller input, when performing a grasp, participants knew they needed to interact with all in-contact buttons; this is comparatively easier than recalling the specific buttons required for pinching.

\section{Limitations and Future Work}

The order of gesture tasks was fixed, with the grasping task always preceding the pinching task. We consciously chose this order based on the anecdotal observation that pinching seemed more difficult in our pilot testing. Although we did not compare gestures directly, this ordering may have introduced learning effects that favored the second (pinching) task, potentially influencing accuracy or user adaptation. Future studies should consider counterbalancing gesture order to control for such effects.

While our subjective questionnaires were designed based on relevant prior work, they were not adapted from validated or standardized instruments. This limits the comparability of our results to other studies using established scales. Future work should incorporate standardized tools for measuring embodiment, usability, and adaptation, such as the Presence Questionnaire or System Usability Scale, to enhance generalizability.

The pinching gesture may have introduced usability challenges. Despite instructions and demonstrations, some participants naturally used pinch variants (e.g., pressing fingers against the palm) that were not detected by the hand tracking system, which required the fingers to be extended. As a result, some pinch attempts were not recognized, potentially affecting both performance and perceived usability. Future research could explore technical limitations in pinch detection methods.

Additionally, a minor software issue was observed by two participants: objects (hammer or ball) could be unintentionally released and lost in the virtual environment if dropped before target contact, slightly increasing completion time. While this bug was reported by only a few participants and likely had minimal impact on overall results, it highlights the need for robust interaction feedback and object constraints in VR task design.

\section{Conclusion}

Our objective was to enhance understanding of varying visual representations across different input modalities. To this end, we conducted a study evaluating combinations of hand and controller avatars across both hand tracking and controller-based input in grasping and pinching tasks commonly used in VR systems (e.g., games). Although a prior similar study~\cite{johnson_exploring_2023} has been conducted, we present a more systematic and thorough exploration of relevant factors by comparing across input type (hand tracking, controller), task (pinch vs.\ grasp), and visual representation (all combinations of hand avatar on/off and controller avatar on/off). Our evaluation included objective performance measurement (i.e., completion data and accuracy) and subjective perception questions. 

Our results confirm previous findings showing faster completion times with controllers compared to hand tracking. However, in terms of accuracy, our findings challenge earlier studies by showing that hand tracking can outperform controllers under certain conditions. Furthermore, although we did not observe a significant interaction between avatar representation and input modality on completion time, we did find a clear effect on accuracy. In particular, matching avatars consistently led to higher accuracy and were more favorably received by participants. Importantly, our results also indicate that the effectiveness of a hand avatar with controllers is task-dependent: accuracy may improve only when the physical gesture aligns well with the visual representation, as seen in grasping tasks, but not in gestures like pinching, where this alignment breaks down.

We note that many VR games (and other non-game VR applications) commonly present avatars that do not match the user’s hand or controller. Our findings thus provide valuable insight to future researchers and designers, offering guidance on determining a suitable visual representation for a given input device across different types of common interactions. Overall, our work contributes to ongoing research by exploring and deepening the understanding of the relationship between input devices and avatar representations and their impact on not only each other but user performance and experience as well.

\bibliographystyle{abbrv}
\bibliography{references}
\end{document}